\documentclass[twocolumn,preprintnumbers,
endnote,nofootinbib,prl]{revtex4}
\usepackage{graphicx}

\usepackage{amsfonts} 
\newcommand{\vev}[1]{\langle {#1} \rangle}
\newcommand{\lsim}{\lesssim}
\newcommand{\gsim}{\gtrsim}

\newcommand{\msol}{M_\odot}
\newcommand{\eq}[1]{Eq.~(\ref{#1})}

\newcommand{\tqcd}{T_{\rm QCD}}

\newcommand{\mpbh}{M_{\rm PBH}}
\newcommand{\LX}{\Lambda_X}

\newcommand{\ord}[1]{\mathcal{O}{(#1)}}
\newcommand{\beq}{\begin{equation}}
\newcommand{\eeq}{\end{equation}}
\newcommand{\bea}{\begin{eqnarray}}
\newcommand{\eea}{\end{eqnarray}}

\newcommand{\vphi}{\varphi}

\begin{document}

\pagestyle{plain}

\title{\boldmath LIGO/Virgo Black Holes from a First Order QCD Phase Transition}

\author{Hooman Davoudiasl
\footnote{email: hooman@bnl.gov}
}

\affiliation{Physics Department, Brookhaven National Laboratory,
Upton, NY 11973, USA}


\begin{abstract}

We propose that $\ord{10\msol}$ black holes observed by LIGO/Virgo originate from a {\it first order} QCD phase transition at a temperature $T_* \lsim 100$~MeV.  This is realized by keeping the quark masses small compared to confinement scale down to $T\sim T_*$,  
making QCD transition first order.  We implement this scenario using a light scalar that could potentially be a good dark matter candidate.

\end{abstract} \maketitle

Direct observations of gravitational waves from mergers of compact stellar objects by the LIGO/Virgo collaborations mark the dawn of a new era in astronomy \cite{Abbott:2016blz}.  An interesting feature of the extant binary merger data is that they apparently point to a population of $\ord{10 \msol}$ black holes \cite{LIGOScientific:2018mvr}, with $\msol \approx 2\times 10^{33}~g$ the Solar mass, which could have an astrophysical origin  
\cite{Belczynski:2009xy,Spera:2015vkd}.  However, apart from the potential 
astronomical information gleaned from these impressive measurements, one is compelled to consider if this new probe can shed light on fundamental questions in particle physics and cosmology.  One of the early attempts soon after the first detection by LIGO was to consider whether the observed merger was of $\sim 30 \msol$ primordial black holes (PBHs) that constitute dark matter \cite{Bird:2016dcv}.  Such PBHs are constrained by various observational data \cite{Niikura:2017zjd} 
and may make up only a fraction $\lsim \ord{10\%}$ of the cosmic dark matter budget \cite{Ali-Haimoud:2016mbv}.    Yet, it is still worthwhile to inquire whether such a population of objects 
can be of primordial origin, though they may not be the main component of dark matter. 

It has long been argued that 
the probability of forming a population of $\sim \msol$ PBHs 
is considerably enhanced during the QCD confining 
phase transition \cite{Jedamzik:1996mr,Jedamzik:1998hc}.  The PBH mass scale is set by the size of the horizon at the time of the transition 
which roughly corresponds to a temperature 
of $\tqcd \approx 160$~MeV \cite{Bhattacharya:2014ara}.  Nonetheless, the typical 
PBH mass in this case is a factor of $\ord{10}$ smaller than that suggested by the LIGO/Virgo data.  
Also, in the Standard Model (SM)  
the QCD phase transition is not first order \cite{Bhattacharya:2014ara}, 
hence not as efficient for the purposes of generating PBHs.  The underlying reason is that during a first order phase transition the speed of sound tends to zero, 
and hence the pressure response of the fluid 
vanishes and does not counter-balance the collapse of horizon-sized primordial over-densities \cite{Schmid:1996qd}.  While 
the pressure response is expected to be lower during the SM QCD transition, the effect 
would not provide the same efficiency as a first order transition. Also, the standard QCD confinement  would  dominantly yield $\ord{\msol}$ 
PBHs \cite{Byrnes:2018clq}.

Arguments based on effective field theories suggest that the QCD transition would be first order 
if the number of light quarks $N_f \geq  3$ \cite{Pisarski:1983ms}, at the onset of confinement.  
The strange quark mass $m_s\approx 100$~MeV 
is not far from the transition temperature $\tqcd$ and 
hence this condition is not satisfied in the SM, in agreement with lattice QCD results \cite{Bhattacharya:2014ara}.  We note that a lattice QCD confirmation of the prediction in 
Ref.~\cite{Pisarski:1983ms} is still under investigation \cite{DElia:2018fjp}.

In this work, we entertain the possibility that the $\ord{10 \msol}$ population of black holes 
points to a first order QCD phase transition, assuming that the number of light quarks $N_f = 6$ 
at the onset of QCD transition.   Given that additional light quarks drive the scale of QCD 
confinement to lower values, one then expects $\tqcd \lsim 160$~MeV in this scenario, corresponding to a lager Hubble volume and hence larger typical PBH masses.

If one arranges for  
the Higgs filed to get a vacuum expectation value (vev) after the QCD transition, at $T\lsim 100$~MeV, one could have $N_f=6$ light quarks during the transition.  Here, quark condensation $\vev{\bar q q}\neq 0$ breaks electroweak symmetry; we dub this symmetry breaking the {\it electro-strong} phase transition.  
This model could result in the desired phase transition, but it is expected to entail a period of supercooling\footnote{
We thank V. Vaskonen for emphasizing this effect, and 
its associated problems, in such a scenario.}.  
In the supercooling phase, the Hubble constant is typically governed by weak scale energies, and thus the formation of $\sim 10 \msol$ does not appear feasible in such a scenario.  While not directly relevant to the subject of this work, as this scenario could potentially yield interesting cosmology, we present a possible model that could lead to Higgs condensation after QCD transition, in the appendix.

To achieve a first order QCD phase transition, leading to $\sim 10 \msol$ PBHs, we hence consider a model in which the dynamics of a light scalar suppresses quark masses before the QCD transition, but results in the measured values afterwards.  This scenario does not entail the above  supercooling, as Higgs condensation takes place at the conventional temperature of $T\sim 100$~GeV.\footnote{The value of the QCD vacuum energy is not precisely known; its estimates cover a fairly broad range \cite{Holdom:2007gg,Narison:2010cg}.  That range could allow for a somewhat lower $\tqcd$, corresponding to a first order QCD transition with $N_f=6$, without leading to supercooling.}  
Generally speaking, we will not address various potential tunings that are required to realize the parameters of the models we will discuss.  Some of these question may be addressed in ultraviolet completions of the effective theories we consider, but that question is beyond the scope of this work.
   
For some alternative ideas on achieving a first order QCD phase transition in the early Universe, see for example, Refs.~\cite{Schwarz:2009ii,Boeckel:2009ej}.  Ref.~\cite{Ipek:2018lhm} considers a first order QCD phase transition at temperatures above the electroweak scale $\sim 100$~GeV, due to a larger initial value of QCD coupling constant.  The possibility of cold baryogenesis from strong CP violation, with delayed electroweak symmetry breaking, was considered 
in Ref.~\cite{Servant:2014bla}.  In Ref.~\cite{Iso:2017uuu}, the possibility of a first order QCD transition with electroweak symmetry breaking after QCD confinement has been considered and analyzed in some detail; see also Re.f~\cite{Arunasalam:2017ajm}
\footnote{Re.f~\cite{Arunasalam:2017ajm} maintains that $\msol$ PBHs form, as it implicitly assumes that no supercooling takes place according to its adopted underlying theory.  Instead, we choose to implement the first order QCD transition through initially suppressed quark Yukawa couplings, assuming electroweak symmetry is broken, thereby avoiding possible  difficulties related to vacuum energy domination.}. 

We will first discuss the 
effect of introducing additional light quarks at QCD confinement transition, from a model-independent point of view.  The mass of PBHs, corresponding to the energy contained within the horizon during the radiation dominated era, can be approximated as \cite{Byrnes:2018clq} 
\beq
M_{\rm PBH}(T) \approx 2.4\, \gamma \,\msol  \sqrt{\frac{61.75}{g_*}} \left(\frac{160\,{\rm MeV}}{T}\right)^2,     
\label{MPBHT}
\eeq
where $\gamma$ is an $\ord{1}$ constant that, depending on the 
amplitude of primordial over-densities, can have values $\sim \text{few}\times (0.1-1)$ \cite{Carr:2009jm,Nakama:2016gzw} and $g_*$ is the relativistic degrees 
of freedom in the primordial plasma.  Thus, 
\eq{MPBHT} suggests that if $\tqcd$ were well below $\sim 160$~MeV then one could take $\mpbh \gg \msol$ to be the typical mass of PBHs formed during the QCD transition.

Next, we address the conditions for achieving a lower confinement temperature.  We will examine how low $\tqcd$ can be if in addition to the SM up and down quarks there are other quarks below $\tqcd$ in the early Universe plasma.  We will focus on the case when all SM quarks 
are light at $\tqcd$, which is the case realized in a model we will propose below.  

For a rough estimate, we first find the value of the QCD coupling 
constant $\alpha_s(\mu_3)$ at $\mu_3 \sim 160$~MeV, corresponding to confinement for the standard $N_f = 3$ case. Then, we will find the scale $\mu_6$, corresponding to 
$N_f= 6$ light quarks, by demanding $\alpha_s (\mu_6)\approx \alpha_s(\mu_3)$; we will use a 
one-loop approximation for the running.  Obviously, this is not meant to be a precision treatment, but only an order of magnitude estimate for the value of $\alpha_s$ that would yield confinement.  

At the one-loop order, we have
\beq
\alpha_s^{-1}(\mu) = \alpha_s^{-1}(\mu_0) + \frac{2N_f - 33}{6 \pi}\, \ln (\mu_0/\mu),
\label{asrun}
\eeq
where $\mu_0$ is a reference scale.  Let us take $\mu_0 = m_Z$, where $m_Z\approx 91.2$~GeV 
is the mass of the $Z$ boson.  We have $\alpha_s(m_Z) \approx 0.118(11)$ \cite{Tanabashi:2018oca}.  It follows from 
\eq{asrun} that 
\beq
\delta \alpha_s = \frac{\delta N_f}{3\pi}\alpha_s^2\, \ln(\mu_0/\mu).
\label{delas}
\eeq
Since in our setup the top quark will be light down to very small temperatures, we estimate that the effect of the extra quark, $\delta N_f=1$, on the value of $\alpha_s(m_Z)$, corresponding to 
running between the top mass and $m_Z$.  Using the above expression, we then find 
$\alpha_s(m_Z)|_{N_f=6} \approx 0.117$, which yields 
$\mu_6 \sim 50$~MeV.  Hence, we may expect QCD with $N_f=6$ to have a confining phase transition scale $\sim 50$~MeV.  At the scales $\mu_3$ and $\mu_6$, we have 
$\alpha_s\sim 5$, corresponding to the onset of confinement.  

Given that 
$
(\mu_3/\mu_6)^2 \sim 10, 
$
at the order of magnitude level, we find  
\beq
\frac{\sqrt{g_*}\,\tqcd^2|_{N_f = 3}}{\sqrt{g_*}\,\tqcd^2|_{N_f=6}} 
\sim \frac{\sqrt{g_*}|_{N_f = 3}}{\sqrt{g_*}|_{N_f=6}}\times 10,
\label{Tratio}
\eeq
where $g_*|_{N_f=3} = 61.75$ for the conventional case, and $g_*|_{N_f=6} = 93.25$ if all SM quarks are light at QCD transition, as could be the case in the discussion presented later on in this work.  Using \eq{MPBHT}, we find  
\beq
\mpbh(\tqcd; N_f=6) \sim 20\, \gamma\, \msol.
\label{MPBHNf6}
\eeq 
The above implies that a ``cooler'' QCD phase transition temperature could be the origin 
of the $\ord{10 \msol}$ black holes observed by LIGO/Virgo.  

To summarize the preceding discussion, we have argued that in the presence of $N_f=6$ light quarks during QCD confinement {\it (i)} a significant enhancement in the efficiency of 
PBH production can be achieved due to a first order phase transition, and {\it (ii)} a boost of the PBH masses by about an order of magnitude to $\ord{10 \msol}$ due to a 
lower transition temperature can result.  

Here, we do not address the relic abundance of the PBHs.  If PBHs comprise a fraction $f$ of the cosmic dark matter budget, some estimates suggest that 
$f\sim 0.001$ \cite{Sasaki:2016jop,Raidal:2017mfl,Ali-Haimoud:2017rtz,Raidal:2018bbj} is needed in order to be consistent with the LIGO/Virgo merger signal.  The value of $f$ depends 
on the probability distribution for the energy density contrast parameter 
$\delta$ (see for example Ref.~\cite{Byrnes:2018clq}).  Our scenario is 
only concerned with the efficiency of producing $\ord{10\msol}$ PBHs; we have implicitly assumed that the required distribution of $\delta$ was realized in the early Universe.
We will next consider models that could in principle implement the above scenario, where $N_f=6$ 
quarks are light at the QCD transition, making it first order.

We aim to present a scenario that would avoid subtleties associated 
with a supercooling period from weak scale vacuum energies.  Here, we assume that 
electroweak symmetry is broken in the usual fashion, at $T\sim 100$~GeV, but  
quarks remain lighter than $\sim 100$~MeV.  To see how this could happen, let right-handed SM 
quarks  $q_R$  and a light scalar field $\varphi$ be odd under a $\mathbb Z_2$.  Then, we need terms of the form 
\beq
\frac{\varphi \, H\,\bar Q_L q_R}{\Lambda_\vphi}\,,
\label{qmop}
\eeq
where $Q_L$ and $H$ are the SM quark and Higgs doublets, respectively.  We assume that $\varphi$ starts at 
$\varphi =0$.  This can be the result of a period of high temperature epoch followed by 
inflation that locks $\vphi$ at the origin.  Phenomenologically, we expect that 
$\Lambda_\vphi \gsim 1$~TeV, given that the SM seems to be a good effective theory up to this energy scale.    

In order for the $\vphi$ to produce the correct top mass, we need the final value of $\vphi \sim \Lambda_\vphi$.  For $\vphi$ to evolve to its final 
value only after QCD confinement, we would generically need $m_\vphi \lsim 10^{-12}$~eV, corresponding to the Hubble scale 
${\cal H}\sim T^2/M_P$ at that era, where the Planck mass $M_P \approx 1.2 \times 10^{19}$~GeV.  For such a light field to avoid causing severe 
deviations from Newtonian gravity, its couplings to nucleons must be very tiny.  However, the largest plausible value 
of the suppression scale in \eq{qmop} is 
$\Lambda_\vphi \sim M_P$.  One would then end up with a 
light field that starts with an initial amplitude of oscillations of $\ord{M_P}$ and a mass of $\sim 10^{-12}$~eV resulting in an energy density $\rho \sim (m_\vphi \, \Lambda_\vphi)^2 \sim 10^{32}$~eV$^4$ at $T\sim \tqcd$, which would be too large for viable dark matter.  Also, the strength of the low energy coupling of $\vphi$ to nucleons will 
end up being $\ord{10^{-20}}$ or so \cite{Knapen:2017xzo}, which is  
too large by a factor of $\sim 10^4$ \cite{Berge:2017ovy}.  To address this issue,   
One could imagine adding various particles at high scales that could lead to cancellations among the effective $\vphi$-gluon couplings.  In what follows, we will sketch such a model.

Let us assume that there is a scalar $\Phi$ of mass $m_\Phi \gsim 100$~GeV and vev $\vev{\Phi}\sim 10\, m_\Phi$.  We also consider a pair of vector-like SM color triplet fermions $F_1$ and $F_2$ and a light scalar $\phi$ with a large initial value.  It is assumed that  
$\mathbb Z_2(\Phi) =\mathbb Z_2(q_R) = -1$.  We can then write down the following interactions [correct $SU(2)_L$ structure implicit]:
\beq
y_q \Phi \bar F_{iL} q_R + \xi_q H \bar Q_L F_{iR} 
+ M_F \bar F_i F_i  + \lambda_i \phi \bar F_i F_i  
+ \text{\small H.C.}\,,
\label{scalar-int}
\eeq
where $i=1,2$ and we have suppressed the flavor index for quarks.  Given the above setup, the following effective operator can be obtained 
\beq
y_q \xi_q\sum_i \frac{\Phi}{\Lambda_{\Phi i}} \, H\,\bar Q_L q_R\,.
\label{Yukawa}
\eeq
The scale $\Lambda_{\Phi i} = M_F + \lambda_i \phi$, where $M_F$ is the $F_{1,2}$  
vector-like mass for $\phi\to 0$, at late times.  The effective $\phi$-dependent Yukawa  
coupling of the SM quarks to the Higgs, 
$\lambda_q H\,\bar Q_L q_R$, is 
then given by
\beq
\lambda_q(\phi) = \sum_i \frac{y_q \xi_q \vev{\Phi}}{M_F + \lambda_i \phi}, 
\label{lamq}
\eeq
where $y_t \xi_t \vev{\Phi}\approx M_F/2$, for we need to recover $\lambda_t (0)\approx 1$ at late times.  We then see, from \eq{lamq}, that $\lambda_q \ll 1$ for $|\lambda_i| \phi \gg M_F$.  

The potential for $\phi$ is simply given by its mass term $(1/2) m_\phi^2 \phi^2$.  Assuming that 
$m_\phi\sim 10^{-12}$~eV, the field $\phi$ will start tracking its potential to $\phi = 0$ after the 
QCD phase transition at $\tqcd \lsim 100$~MeV.  If the initial $\phi$ energy density  
$\rho(\tqcd) \sim m_\phi^2 \phi^2 \sim 10^{24}$~eV$^4$, by the time of matter-radiation equality at $T_{mr}\sim 1$~eV in standard cosmology, $\rho$ 
will be diluted by $(T_{mr}/\tqcd)^3 \sim 10^{-24}$ to achieve the standard value 
$\rho(T_{mr})\sim \text{eV}^4$.  Hence, $\phi$ could be a viable dark matter candidate 
for $\phi_i \sim 10^{15}$~GeV, where $\phi_i$ is the initial value of $\phi$.

As before, the above setup would lead to severe deviations from Newtonian gravity if $\phi$-nucleon coupling $y\gsim 10^{-24}$ \cite{Berge:2017ovy}.  This could, for example, be mediated 
by top quark mixing with $F_{1,2}$ from the interactions (\ref{scalar-int}).  We want $M_F$ to be small compared to $|\lambda_i| \phi$ and $y_t \xi_t \vev{\Phi} \approx M_F/2$.  On the other hand, $\phi\lsim 10^{15}$~GeV from the above discussion, so that $\phi$ oscillations do not overclose the Universe.  Hence, we conclude that the typical $\phi$-nucleon coupling would be too large, unless 
there is a cancellation among various contributions.  We find that if the couplings $y_q$ and 
$\xi_q$ are equal, while $\lambda_1 + \lambda_2 = 0$, then the contributions of loop diagrams mediated by $F_1$ and $F_2$ that induce $\phi$-nucleon coupling yield $y_n=0$ today, corresponding to $\phi=0$ (see, for example, Ref.\cite{Gubser:2004du} for a possible implementation of such interactions in a string theory context).  One could have a sufficiently small $y_n$ if the model parameters have only tiny deviations from the above assumed values.

Let $\lambda_1 = -\lambda_2 = \lambda$.  
We then find that $\lambda_q (\phi_i)\sim - y_q \xi_q \vev{\Phi} M_F/(\lambda \phi_i)^2$.  
To have light top quarks during the QCD transition, 
we require that $\lambda_t(\phi) \lsim 10^{-3}$.  If the $\phi$ condensate is a viable dark matter candidate, then we obtain $M_F/\lambda \lsim 5 \times 10^{13}$~GeV. 

We then conclude that the above model provides a reasonable picture for how the QCD phase transition could be first order, at a somewhat lower temperature, while providing 
a possible dark matter candidate from $\phi$ oscillations.  We note that the required mass for 
$\phi$ can be naturally obtained after the first order QCD transition if there is a Planck scale coupling $\sim \phi^2\, \text{Tr} [G_{\mu\nu}G^{\mu\nu}]/M_P^2$, where 
$G_{\mu\nu}$ is the gluon field strength tensor.  The value of the gluon condensate is 
estimated to be $\vev{\alpha_s GG}\sim 10^{-2}$~GeV$^4$ \cite{Shifman:1978by,Shifman:1978bx,Gubler:2018ctz}, with $\alpha_s$ the strong coupling constant.  As in our setup QCD confines at a lower scale, we estimate $\vev{\alpha_s GG}\sim 10^{-3}$~GeV$^4$ for our model with $N_f=6$ light quarks, which yields the right order of magnitude for $m_\phi$.  

Here we would like to describe some of the potential phenomenological 
predictions of the above scenarios.  
Assuming that the above setup can realize a first order QCD transition, motivated by the  
masses of the black holes observed by LIGO/Virgo, we can expect primordial gravitational 
waves corresponding to an epoch $\tqcd\sim 50-100$~MeV.  Such signals may be detectable 
using pulsar timing arrays.  This idea was first discussed in Ref.~\cite{Witten:1984rs} and 
more recently studied in Ref.~\cite{Caprini:2010xv}; see also Ref.~\cite{Schwaller:2015tja}.

If $\phi$ is not identified as dark matter, one could possibly consider values of $M_F\sim$~TeV.  In that case, the degenerate fermions $F_i$ may be within the reach of the LHC. From 
the interactions (\ref{scalar-int}) one could expect their main decay channels to be 
$F_i \to H t$, $F_i \to \Phi t$, $F_i \to Z t$, and $F_i \to W b$.  If $m_\Phi > 2m_t$, a possible 
decay channel for $\Phi$ would be into a pair of top quarks. 

To summarize, motivated by the $\ord{10\msol}$ black holes 
observed by LIGO/Virgo in binary mergers, 
we entertained the possibility that the QCD phase transition was first order due to the effect of 
6 light quarks.  The larger number of light quarks, compared to the standard case, 
pushes the transition temperature below $\sim 100$~MeV.  The first order nature of the transition 
significantly improves the likelihood of forming primordial black holes and its lower temperature 
suggests that these black holes 
can potentially be as heavy as $\sim 10 \msol$, compared to $\sim 
\msol$ for the standard QCD transition.  We presented a model that could potentially realize the above scenario and yield a good dark matter candidate, which is a light scalar of mass $\sim 10^{-12}$~eV.


\acknowledgments

We thank S. Dawson, P. Petreczky, R. Pisarski, and P. Serpico, for discussions and comments.  The author is grateful to D. Morrissey and V. Vaskonen for helpful comments on earlier versions of the manuscript.  This work is supported by the United States Department of Energy under Grant Contract DE-SC0012704.


\appendix*

\section{Higgs Condensation after QCD Transition}

Let  $\sigma$ be a scalar 
that starts out massless which is stuck at $\sigma = \sigma_i$ due to Hubble friction.  We assume that $\sigma$ controls the coupling constant $g_X$ of a new confining gauge interaction $X$, {\it e.g.} an $SU(N)$ gauge sector:
\beq
-\frac{1}{4\, g_X^2} \text{Tr} [X_{\mu\nu}X^{\mu\nu}] 
-\frac{\sigma^2}{M_X^2} \text{Tr} [X_{\mu\nu}X^{\mu\nu}] \,,
\label{X}
\eeq
where we require $\sigma_i \sim M_X$.  The effective value of the coupling 
$g_{{\rm eff} X}$ is given 
by 
\beq
g_{{\rm eff}X}^{-2} = g_X^{-2} + 4 \sigma^2 M_X^{-2}.
\label{gXbar}
\eeq
Therefore, $g_{{\rm eff} X}$ is in the weak coupling regime as long as $\sigma_i \sim M_X$.  Let  
$\sigma$ couple to the SM $SU(3)_c$ color 
\beq
-\frac{1}{4\, g_s^2} \text{Tr} [G_{\mu\nu}G^{\mu\nu}] 
-\frac{\sigma^2}{M_s^2} \text{Tr} [G_{\mu\nu}G^{\mu\nu}] \,,
\label{GG}
\eeq 
where $G_{\mu\nu}$ is the gluon field strength.  The above interaction will endow $\sigma$ with an initial mass $m_{\sigma i}^2 \sim \vev{GG}/M_s^2$, 
upon color confinement.  We assume   
$M_s \gg M_X$ so that $\sigma_i \sim M_X$ allows gluons to confine  
at $T\lsim 100$~MeV.  

Finally, the Higgs potential is taken to have the form  
\beq
V_H = \left[-\frac{\text{Tr} [X_{\mu\nu}X^{\mu\nu}]}{M_V^2} + m_H^2\right] H^\dagger H + \lambda (H^\dagger H)^2\,,
\label{VH0}
\eeq
where $M_V$ is a large mass scale connecting the SM with the new gauge sector (for other 
applications of such connections, see for example Ref.~\cite{Davoudiasl:2017zws}).  We will denote the $X$ gauge sector confinement scale by $\Lambda_X \gg m_H$, where $m_H^2\geq 0$ is the initial Higgs mass that is sufficiently small to allow Higgs condensation\footnote{We thank D. Morrissey for comments on the Higgs potential.}.  
By analogy with QCD, 
$\vev{XX}\sim \LX^4$, resulting in the Higgs mass parameter 
\beq
\mu_H^2 \sim - \LX^4 M_V^{-2} + m_H^2 \sim - (100)^2~\text{GeV}^2\,,
\label{muH2}
\eeq
resulting in the Higgs boson mass $M_H\approx 125$~GeV \cite{Tanabashi:2018oca}.

We want $\sigma\to 0$, leading to $X$ 
confinement and electroweak symmetry breaking, 
after QCD transition.  Thus results in a period of supercooling 
governed by ${\cal H}_{sc} \sim \LX^2/M_P$.  A sensible effective theory demands 
$M_X \gsim\LX\gsim |\mu_H|$; we may set $\LX \sim 1$~TeV, which gives $M_V\sim 10$~TeV and ${\cal H}_{sc} \sim 10^{-4}$~eV.  
Non-perturbative QCD effects can endow $\phi$ with a mass $m_{\sigma i} \gsim 3 {\cal  H}$, in order for $\sigma$ to roll to the origin.  For standard QCD, $\vev{\alpha_s \,GG}\sim 10^{-2}$~GeV$^4$  \cite{Shifman:1978by,Shifman:1978bx,Gubler:2018ctz}.  Due to $N_f=6$ massless quarks the QCD confinement scale is lower and we estimate $\vev{\alpha_s\, GG}_{N_f=6}\sim 10^{-3}$~GeV$^4$.  \eq{GG} then yields $m_{\sigma i} \gsim 10^{-4}$~eV for $M_s \lsim 10^{12}$~GeV.  The initial energy density 
$\rho_\sigma \sim m_{\sigma i} m_{\sigma f} M_X^2 \gsim 
\LX^4 M_X/M_P \gsim 10^{27}$~eV$^4$ will be somewhat suppressed at the end of  the supercooling phase.  If this suppression is not sufficient, one must assume that $\sigma$ can decay 
before BBN. 

To avoid thermalizing $\sigma$, we require that $X$ production is inefficient: 
$T_{\rm RH}^5/M_V^4 \ll {\cal H}_{sc}$, where $T_{\rm RH}$ is the initial reheat 
temperature.  For $T_{\rm RH}\sim 1$~GeV as a possible value, 
we find $M_V\gsim \text{few}\times 1$~TeV, consistent with the above reference value 
$M_V\sim 10$~TeV.  Using \eq{X}, one finds  
$m_{\sigma f}^2 \sim \LX^4/M_X^2$.  If  
$m_H \gg T_{\rm RH}$, an appreciable Higgs population will not be produced.

An interesting aspect of the {\it electro-strong} phase transition is 
that electroweak vector bosons remain 
massless down to $\tqcd\lsim 100$~MeV, instead of the usual 
$T\sim 100$~GeV.  Given the significance of sphalerons for models of baryogenesis, one could 
end up with novel possibilities to realize the baryon asymmetry of the Universe (see, {e.g.}, 
Ref.~\cite{Servant:2014bla}).

The coupling of a hidden gauge sector to the Higgs may allow for potential signals at the LHC.  This was studied in some detail in Ref.~\cite{Craig:2015pha}, though for values of parameters different 
from those chosen in the our discussion (for earlier work, see also 
Refs.\cite{Strassler:2006im,Juknevich:2009ji,Juknevich:2009gg}).  Nonetheless, the above model or its modifications 
could lead to possible detectable signals through Higgs-mediated production of the $X$ sector 
hadrons.  Making such determinations would require a more detailed analysis in future work.  For potential cosmological consequences of ``$X$ glueballs,'' see for example Refs.~\cite{Soni:2016gzf,Forestell:2016qhc}.
  

\end{document}